\documentclass[prl,aps,twocolumn,superscriptaddress,showpacs]{revtex4}
\usepackage{graphicx}

\begin{document}

\title{The Optimal Faucet}

\author{ H. Henry Chen}
\affiliation{Department of Physics,
Harvard University, Cambridge, MA 02138}
\email{chen@physics.harvard.edu}

\author{Michael P. Brenner}
\affiliation{Division of Engineering and Applied Sciences,
Harvard University, Cambridge, MA 02138}
\email{brenner@deas.harvard.edu}

\date{\today}

\begin{abstract}
The production of small fluid droplets relies on an instability of
solutions to the Young-Laplace equation. We ask whether smaller
droplets can be produced by changing the shape of the nozzle. At a
given critical pressure, the circular nozzle actually produces the
largest droplet. The droplet volume can be decreased by up to 18\%
using a triangular nozzle with stretched corners.
\end{abstract}


\maketitle

A standard protocol for producing small droplets is as follows:  a
pipette, of circular cross-section, is pressurized at one end, pushing
out a small fluid droplet. If the nozzle is sufficiently small, force
balance requires that the droplet has constant mean curvature. At a
critical pressure, this equilibrium shape becomes unstable, ultimately
leading to the droplet detaching from the nozzle.

The volume of fluid entrained during this process is set by the total
fluid volume contained in the critical droplet. This volume scales
like $r^3$, where $r$ is the nozzle radius. On the other hand, the
critical pressure for ejecting this droplet scales like $\gamma/r$,
where $\gamma$ is the liquid surface tension. Thus, ejecting smaller
droplets requires higher pressures.  The smallest size
droplet that can be ejected is thus determined by the highest pressure
that can be reliably applied to the nozzle, without material failure,
etc.

One strategy for creating smaller droplets than those dictated by the
instability of a static droplet is to use a time varying forcing at
the nozzle.  This method has achieved an order of magnitude decrease
in droplet volume \cite{basaran02}.

However, typical nozzles use a circular cross section. It is not
unreasonable to imagine that changing the shape of the cross
section to be some other shape may decrease the ejected droplet
volume, while maintaining the same applied pressure. For example,
imagine that we have a circular nozzle with a pendant droplet just
below the critical volume: by ``squeezing'' the shape of the
nozzle cross section into an elliptical shape, one might cause the
droplet to detach at a lower volume.

In this paper we address the question: what is the shape of a
nozzle for which the ejected droplet volume is minimized, for a
given applied pressure? We demonstrate that circular nozzles do
not eject the smallest droplets; instead, the optimal nozzle more
closely resembles an equilateral triangle, albeit with
``stretched'' corners. The best nozzle shape that we have found
has an ejected droplet volume about twenty percent smaller than
the circular nozzle with the same critical pressure.  Our method
is inspired by and extends J. Keller's classic treatment of the
Euler buckling problem with a beam of nonuniform cross section
\cite{keller60}.  Recently, the method has been applied to the
optimization of a bistable switch \cite{brenner03}.  For a detailed
mathematical treatment of capillary surfaces in general, see \cite{finn}.

This Letter is organized as follows.  We first explain the origin
of the pendant droplet instability.  Then we describe our method
for reducing droplet size.  Lastly we provide numerical
calculations implementing the method, and present the candidate
optimal nozzle.

\textit{Pendant Droplet Instability.} - The instability of a
droplet protruding from a nozzle is due to a bifurcation, most
easily seen in the case of a circular nozzle that is much smaller
than the capillary length, which allows us to neglect gravity. The
shape of the droplet is then determined by the Young-Laplace
equation $p = \gamma K$, where $p$ is the pressure difference
across the liquid/air interface, $\gamma$ is the surface tension,
and $K$ is the mean curvature of the droplet surface. This
equation describes a surface of constant mean curvature $p/\gamma$
with the nozzle edge as its boundary.  If the boundary is a
circle, then the solution must be a section of the sphere with
mean curvature $p/\gamma$. From the familiar relation
\begin{equation} K_{\textrm{sphere}} = \frac{2}{\textrm{sphere
radius}}
\end{equation}
we deduce that the radius of curvature of
the droplet is $2\gamma/p$.  For small $p$, such that the sphere
radius is much greater than the nozzle radius, the solution is a
shallow spherical cap.  But note that its complement, the rest of
the sphere, is also a solution. As $p$ is increased, these two
solutions approach each other until both become a hemisphere
with the nozzle at the equator.
The pressure at which the two solutions meet is the critical pressure
$p^*$, and the corresponding degenerate solution is unstable.  Note that
the critical pressure is also the maximum pressure, for the nozzle cannot
support a sphere smaller than itself.

For a noncircular nozzle, we no longer have such a simple
geometric picture, however key features remain.  The unstable
solution is still characterized by a bifurcation at which two
solutions meet, corresponding to the maximum pressure achievable
for the given nozzle. The critical pressure for a general nozzle
can be computed as follows:  let the droplet surface be
parameterized as a function $\vec{R}(u,v)$ over a domain $D$ in 
the $uv$-plane, which takes value in three dimensional
physical space.  The boundary of the domain $\partial D$ corresponds
to a closed curve $C$ which represents the nozzle. The curvature
is a nonlinear functional of the surface and its derivatives up to
second order, hence the equation for the droplet shape has the
form
\begin{equation}\label{yl}
\gamma K[\vec{R},\vec{\nabla}\vec{R},\vec{\nabla}\vec{\nabla}\vec{R}]=p,
\end{equation}
where $\vec{\nabla}$ is the gradient operator in the $uv$-plane.

Upon increasing the pressure $p\to p + \delta p$, the surface
changes: $\vec{R} \to \vec{R}+ \delta\vec{R}$. Equation (\ref{yl})
implies that the variation $\delta\vec{R}$ and $\delta p$ are
related by
\begin{equation}\label{variation}
\gamma \hat{L}\delta\vec{R} = \delta p,
\end{equation}
where $\hat{L}\delta\vec{R}$ is the change in mean curvature
induced by the surface change.  $\hat{L}$ is a differential
operator acting on $\delta \vec{R}$.

At the critical solution, the pressure is at a maximum; therefore,
there must be a solution $w=\delta \vec R$ to equation
(\ref{variation}) with $\delta p=0$. The solution $w$ satisfies
\begin{equation}\label{bifurcation}
\hat{L}w = 0
\end{equation}
with boundary condition $w = 0$ at $\partial D$.   Note that the
pressure dependence in this formula arises because
$\hat{L}=\hat{L}[\vec{R}]$ depends implicitly on the pressure $p$
through $\vec{R}$. Hence, the existence of a nonzero $w$ is a
diagnostic for finding the  critical solution to (\ref{yl}) and
the corresponding critical pressure $p^*$.

\textit{Optimization Method.} - Now, to find the optimal nozzle,
we need to derive a relation between the change in critical
pressure and change in nozzle shape.  Since pressure and
volume are conjugate variables, increasing critical pressure is tantamount
to decreasing critical volume.  By iteratively changing the nozzle shape
to increase critical pressure, we will thus arrive at a nozzle which produces
smaller droplets.
We compare the critical volume of the deformed nozzle with that of the
circular nozzle that corresponds to the \textit{same critical pressure},
since pressure is the control variable in practical situations.

Suppose that a given nozzle
shape $C$ has a critical pressure $p^*$, a critical droplet shape
$\vec{R}^*$, and a corresponding $w$. All of these quantities
change when the nozzle shape $C\to C+\delta C$. The change in the
droplet shape $\delta \vec{R}$ is linearly related to the pressure
change $\delta p$  by equation (\ref{variation}) with the boundary
condition $\delta\vec{R}=\delta C$ at $\partial D$. On the other
hand, since the critical solution maximizes the critical pressure,
 $w$ does not change to leading order in $\delta C$.

The change in \textit{critical pressure} induced by $\delta C$ can
therefore be computed by taking the inner product of both sides of
(\ref{variation}) with $w$:
\begin{eqnarray*}
\gamma \langle w, \hat{L}\delta R \rangle
    &=& \gamma \langle \delta R, \hat{L}w \rangle + \gamma \oint b(\delta R,w)\\
  &=& 0 + \gamma \oint b(\delta C,w)\\
  &=& \langle w,\delta p \rangle.
\end{eqnarray*}
Therefore
\begin{equation}\label{boundary}
\delta p = \frac{\gamma \oint b(\delta C,w)}{\langle w,1 \rangle}.
\end{equation}
Here $b(\bullet,\bullet)$ denotes the boundary integrand from
integrating by parts. The derivation also uses the self
adjointness of $\hat{L}$, which is readily demonstrable by
explicit computation \footnote{ In computing an inner product
$\langle f, \hat{L} g \rangle \equiv \int du dv f \hat{L} g,$
where $f$ and $g$ are arbitrary functions of $u,v$, and the
integration is over $D$, we can undo the differentiation on $g$,
and instead let the adjoint operator $\hat{L}^\dag$ act on $f$.
Self-adjointness ($\hat{L} = \hat{L}^\dag$)  can be verified
explicitly. In the derivation, this allows us to simply
interchange $f$ and $g$, while introducing the boundary terms from
integration by parts.}. Equation (\ref{boundary}) is an explicit
relation between a change in the nozzle shape ($\delta C$) and the
resulting change in critical pressure.

\textit{Explicit Formula for $\delta p$.} - We choose the nozzle
$C$ to lie in the $xy$-plane, enclosing the origin. Then the
droplet surface may be given in spherical coordinates by the
distance from the origin ($R$) as a function of the two angles
$\theta \in [0,\pi/2]$ and $\phi \in (0, 2\pi]$.  To avoid the
coordinate singularity at the pole ($\theta = 0$) we use $u,v$
given by $u = \tan(\theta/2)\cos(\phi)$ and $v =
\tan(\theta/2)\sin(\phi)$. Hence the surface is a scalar function
$R(u,v)$; its domain $D$ is the unit disk in the $uv$-plane.  We
retain $\phi$ to denote the polar angle in the $uv$-plane.

An appealing feature of this coordinate system is that the line element
remains diagonal:
\[ds^2 = dR^2 + \Gamma (du^2+dv^2),\]
where $\Gamma = 4R^2/(1+u^2+v^2)^2$.  It is then straightforward to compute
the free energy
$E = \int (\gamma dA - p dV)$
which yields, upon variation, the Young-Laplace equation
\begin{equation}\label{pde}
-\vec\nabla \cdot (C \vec\nabla R) + A R = F,
\end{equation}
where $\vec\nabla$ is the usual gradient operator in the $uv$-plane.
The coefficients are
\begin{eqnarray*}
C &=& \frac{1}{\sqrt{1+(\frac{1+\rho^2}{2R})^2 (\vec\nabla R)^2}},\\
A &=& C \left(\frac{(\vec\nabla R)^2}{R^2} + \frac{8}{(1+\rho^2)^2}
      \right),\\
F &=& p \frac{4R^2}{(1+\rho^2)^2},
\end{eqnarray*}
where $\rho^2 \equiv u^2 + v^2$ is the radial coordinate in the
$uv$-plane.  $\partial D$ corresponds to $\rho = 1$.

In our coordinate system, the pressure change is
\begin{equation}\label{dp}
\delta p = \frac{1}{\delta_w V}
           \oint d\phi \,\delta C \frac{w_\rho R (R^2+R_\phi^2)}
             {(R^2+R_\rho^2+R_\phi^2)^{3/2}},
\end{equation}
where $\delta_w V \equiv \int d^2\rho \,\, w \frac{4R^2}{(1+\rho^2)^2}$.
Here and in the following we use subscripts to denote partial
differentiation.

We can recast this expression into a form that is more geometric.
First, the contact angle $\alpha$ between the drop and the plane of
the nozzle is given by $\cot \alpha(\phi) =
R_\rho/(R^2+R_\phi^2)^{1/2}|_{\rho=1}$ where the right hand side is
evaluated at the boundary.  Second, we define $w_\perp \equiv
w_\rho/(R^2+R_\phi^2)^{1/2}|_{\rho=1}$ which can be understood as
follows - note that $w$ is the difference between the outer and
inner solutions as the pressure approaches bifurcation.  Using the
contact angle given above, this expression is the difference between
the slopes (with respect to the vertical) of the outer and inner
solutions at the boundary.  This is a coordinate independent quantity.
Third, we observe that
\[d\phi \, \delta c \, R = \left(d\phi \sqrt{R^2+R_\phi^2}\right)\left(\delta c
\frac{R}{\sqrt{R^2+R_\phi^2}}\right) = dl\,\delta N,\] where $dl$ is
the line element, and $\delta N$ is the change of the nozzle in the
direction locally normal to the nozzle.  Lastly, the denominator $\delta_w V$ in
(\ref{dp}) is just the change in volume from changing the surface by
$w$.  Putting these facts together, the pressure change
is
\[\delta p = \frac{1}{\delta_w V} \oint dl\,\delta N \,w_\perp\,
\sin^3\alpha,\] which leads to the prescription for changing the nozzle
\begin{equation}\label{iterate}
\delta N \sim \frac{1}{\delta_w V} \,w_\perp\, \sin^3\alpha.
\end{equation}
Clearly, for the circular nozzle, symmetry implies that $\delta N$
should be constant.  But this amounts to a mere reduction in the size of the nozzle; the shape remains a circle.  So the circular nozzle is at an extremum, in fact a minimum of critical pressure for \textit{fixed nozzle area}.

For a noncircular nozzle, the contact angle isn't constant, and hence
the change according to the above formula cannot be constant.  So one
may change the critical pressure while fixing the nozzle area.
Moreover, since the circular nozzle is the only one (except the
infinite strip) with a constant contact angle, the process of
deformation does not end.

We apply (\ref{iterate}) iteratively to a perturbed circular nozzle to
see how the shape evolves away from the circle.
\begin{figure}\begin{center}
\includegraphics[width=3.4in]{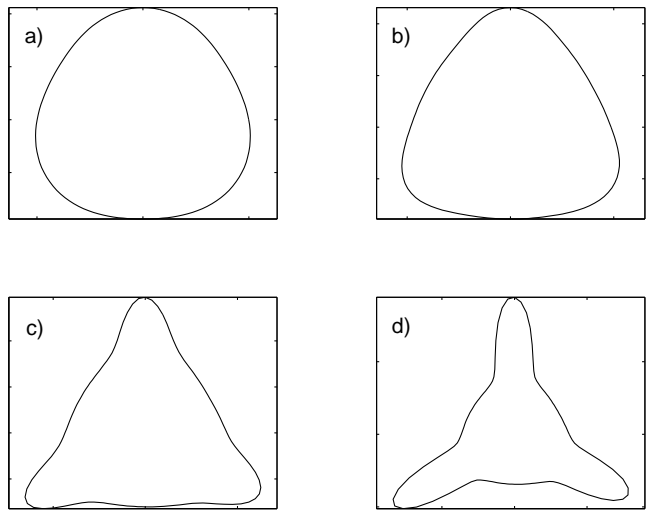}
\caption{\label{evolve3}Evolution of nozzle shape with threefold
symmetry. a) Initial nozzle: $\tilde{V}=1.00$; b) $\tilde{V}=0.97$; c)
$\tilde{V}=0.88$; d) $\tilde{V}=0.82$. $\tilde{V}$ is a normalized
volume given by (\ref{volume}).}
\end{center}\end{figure}
Figure \ref{evolve3} shows the result of iterations starting with a
circle deformed by a perturbation with a three-fold symmetry.  The
perturbation grows with each iteration, and eventually the nozzle
shape becomes concave.  With each iteration, we have applied a
rescaling in order to maintain the nozzle area.  Without the area
constraint, the nozzle would become arbitrarily small in accordance
with (\ref{iterate}).  We are interested in the shape of the nozzle,
not its size.  We also apply the Savitzky-Golay filter \cite{press} at
each iteration to smooth out the mesh noise.  The solutions to the
Young-Laplace equations are obtained using the nonlinear PDE solver in
the MATLAB$^\textrm\textregistered$ PDE Toolbox, which implements the
finite element method for elliptic equations with variable
coefficients, exactly of the form in (\ref{pde}).  For each nozzle
shape, we start at a pressure below the bifurcation and by choosing
different trial solutions obtain both solutions.  Then we bring both
solutions to just below the critical pressure by stepping up the
pressure, using the solution at each step as the trial solution for
the next step.  We then use the average of the two solutions for our
surface, and their difference for $w$.  The validity of this procedure
can be rigorously shown for a circular nozzle, and we expect it to
remaind valid for noncircular nozzles as long as the pressure is brought close
to critical.

In order to compare and select among nozzle shapes, we need a
measure of optimality independent of size.  For every nozzle, we
rescale its critical volume by the critical volume corresponding
to the circular nozzle with the \textit{same} critical pressure.
This dimensionless volume is given by
\begin{equation}\label{volume}
\tilde{V} = \frac{v^*}{\frac{2\pi}{3} \left(\frac{2}{p^*}\right)^3}.
\end{equation}

\begin{figure}\begin{center}
\includegraphics[width=3.4in]{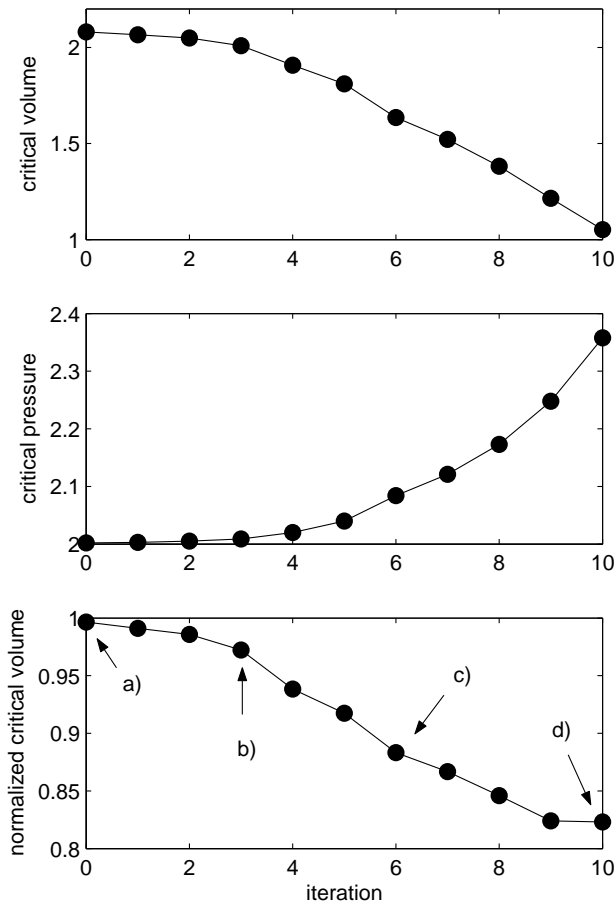}
\caption{\label{iteration}Sequence of iterations away from the circular
nozzle with an initial three-fold perturbation.  The normalized critical
volume given by (\ref{volume}) is shown in the bottom graph.
The arrows indicate the corresponding shapes in Figure \ref{evolve3}.}
\end{center}\end{figure}

Figure \ref{iteration} shows a particular sequence of critical
properties obtained through our iteration procedure. We see that
the critical pressure begins to increase rapidly about the fifth
iteration, after which the decrease in $\tilde{V}$ slows down, and
the nozzle shape becomes stretched out (see Figure \ref{evolve3}).
This means that in order to decrease droplet size at a given
pressure, one should use a nozzle shape that is roughly
triangular, perhaps with somewhat stretched out corners; but
further deformation does not lead to significant improvement.
Moreover, gravitational instabilities will inevitably become
relevant if the ``arms'' become too long \cite{pitts74,michael76}.

It should be emphasized that we have shown a \textit{particular} example of
an improved nozzle, generated by a choice of the initial perturbation.
We have tried other perturbations, leading to shapes with, say, four-fold
symmetry or without any symmetry,
but the three-fold perturbation has yielded the biggest reduction
in the normalized critical volume.

%conclusion
So far we have ignored the effects of gravity, but our formalism
applies just as well to the problem with gravity.  Including gravity
means that the pressure would no longer be constant throughout the
drop surface, but rather a linear function of height: $p \to p -\rho_m
g h(u,v)$, where $\rho_m$ is the mass density of the liquid, $g$ is 
the gravitational acceleration, $h$ is the distance below the nozzle, 
and $p$ now denotes the pressure at the nozzle ($h = 0$).
Although (\ref{pde}) acquires a new term as a result, this term does
not contain derivatives and thus does not contribute to the boundary
integral.  So our formula for the pressure change remains the same in
the presence of gravity.  To be sure, the nozzle evolution would
differ because the contact angle and $w_\perp$ will be affected by
gravity.  Moreover, if the nozzle is too large relative to the
capillary length, then gravity destabilizes all solutions: it is not
possible to suspend a water drop from a meter wide faucet.  It would
be interesting to examine the case of the intermediate sized nozzle,
small enough to have stable solutions, yet large enough to be affected
by gravity.

We thank Eric Lauga and Daniel Podolsky for useful discussions at
various stages of the work. This research is funded by NSF
DMS-0305873 and Harvard MRSEC.

\bibliography{faucet}

\end{document}